\documentclass[%
 reprint,
superscriptaddress,
 amsmath,amssymb,
 aps,
 prl,
 showpacs
]{revtex4-1}
\usepackage{subfigure}
\usepackage{graphicx}
\usepackage{dcolumn}
\usepackage{bm}
\usepackage[final]{pdfpages}
\begin{document}
\title{Accelerator measurements of magnetically-induced radio emission from particle cascades with applications to cosmic-ray air showers}
\author{K.~Belov}
\affiliation{Dept. of Physics and Astronomy, Univ. of California, Los Angeles, Los Angeles CA 90095, USA}
\affiliation{Jet Propulsion Laboratory, Pasadena CA 91109, USA}
\author{K.~Mulrey}
\affiliation{Dept. of Physics and Astronomy, Univ. of Delaware, Newark DE 19716, USA}
\author{A.~Romero-Wolf}
\affiliation{Jet Propulsion Laboratory, Pasadena CA 91109, USA}
\author{S.~A.~Wissel}
\email[Corresponding author: ]{swissel@physics.ucla.edu, swissel@calpoly.edu}
\affiliation{Dept. of Physics and Astronomy, Univ. of California, Los Angeles, Los Angeles CA 90095, USA}
\affiliation{Physics Dept., California Polytechnic State Univ., San Luis Obispo CA 93407, USA}
\author{A.~Zilles}
\affiliation{Institut f\"{u}r Experimentelle Kernphysik, Karlsruher Institut f\"ur Technologie, 76128 Karlsruhe, Germany}
\author{K.~Bechtol}
\affiliation{Kavli Institute for Cosmological Physics, Univ. of Chicago, Chicago IL 60637, USA}
\author{K.~Borch}
\affiliation{Dept. of Physics and Astronomy, Univ. of California, Los Angeles, Los Angeles CA 90095, USA}
\author{P.~Chen}
\affiliation{Dept. of Physics, Graduate Institute of Astrophysics, Leung Center for Cosmology and Particle Astrophysics, National Taiwan University, Taipei 10617, Taiwan}
\author{J.~Clem}
\affiliation{Dept. of Physics and Astronomy, Univ. of Delaware, Newark DE 19716, USA}
\author{P.~W.~Gorham}
\affiliation{Dept. of Physics and Astronomy, Univ. of Hawaii, Manoa HI 96822, USA}
\author{C.~Hast}
\affiliation{SLAC National Accelerator Laboratory, Menlo Park CA, 94025, USA}
\author{T.~Huege}
\affiliation{Institut f\"ur Kernphysik, Karlsruher Institut f\"ur Technologie, 76021 Karlsruhe, Germany}
\author{R.~Hyneman}
\affiliation{Dept. of Physics and Astronomy, Univ. of California, Los Angeles, Los Angeles CA 90095, USA}
\affiliation{Physics Dept., College of William \& Mary, Williamsburg VA 23187, USA}
\author{K.~Jobe}
\affiliation{SLAC National Accelerator Laboratory, Menlo Park CA, 94025, USA}
\author{K.~Kuwatani}
\affiliation{Dept. of Physics and Astronomy, Univ. of California, Los Angeles, Los Angeles CA 90095, USA}
\author{J.~Lam}
\affiliation{Dept. of Physics and Astronomy, Univ. of California, Los Angeles, Los Angeles CA 90095, USA}
\author{T.~C.~Liu}
\affiliation{Dept. of Physics, Graduate Institute of Astrophysics, Leung Center for Cosmology and Particle Astrophysics, National Taiwan University, Taipei 10617, Taiwan}
\author{J.~Nam}
\affiliation{Dept. of Physics, Graduate Institute of Astrophysics, Leung Center for Cosmology and Particle Astrophysics, National Taiwan University, Taipei 10617, Taiwan}
\author{C.~Naudet}
\affiliation{Jet Propulsion Laboratory, Pasadena CA 91109, USA}
\author{R.~J.~Nichol}
\affiliation{Dept. of Physics. and Astronomy, University College London, London WC1E 6BT, United Kingdom}
\author{B.~F.~Rauch}
\affiliation{Dept. of Physics, Washington Univ. in St. Louis, St. Louis MO 63130, USA}
\author{B.~Rotter}
\affiliation{Dept. of Physics and Astronomy, Univ. of Hawaii, Manoa HI 96822, USA}
\author{D.~Saltzberg}
\affiliation{Dept. of Physics and Astronomy, Univ. of California, Los Angeles, Los Angeles CA 90095, USA}
\author{H.~Schoorlemmer}
\affiliation{Dept. of Physics and Astronomy, Univ. of Hawaii, Manoa HI 96822, USA}
\author{D.~Seckel}
\affiliation{Dept. of Physics and Astronomy, Univ. of Delaware, Newark DE 19716, USA}
\author{B.~Strutt}
\affiliation{Dept. of Physics. and Astronomy, University College London, London WC1E 6BT, United Kingdom}
\author{A.~G.~Vieregg}
\affiliation{Kavli Institute for Cosmological Physics, Univ. of Chicago, Chicago IL 60637, USA}
\affiliation{Dept. of Physics, Enrico Fermi Institute, Univ. of Chicago, Chicago IL 60637, USA}
\author{C.~Williams}
\affiliation{Dept. of Physics, Stanford Univ., Stanford CA, 94305, USA}
\collaboration{T-510 Collaboration}\noaffiliation
\date{\today}
\begin{abstract}
For fifty years, cosmic-ray air showers have been detected by their radio emission.  We present the first laboratory measurements that validate electrodynamics simulations used in air shower modeling. An experiment at SLAC provides a beam test of radio-frequency (RF) radiation from charged particle cascades in the presence of a magnetic field, a model system of a cosmic-ray air shower. This experiment provides a suite of controlled laboratory measurements to compare to particle-level simulations of RF emission, which are relied upon in ultra-high-energy cosmic-ray air shower detection. We compare simulations to data for intensity, linearity with magnetic field, angular distribution, polarization, and spectral content. In particular, we confirm modern predictions that the magnetically induced emission in a dielectric forms a cone that peaks at the Cherenkov angle and show that the simulations reproduce the data within systematic uncertainties.
\begin{description}
\item[PACS numbers] 
95.55.Vj, 98.70.Sa, 29.27.-a
\end{description}
\end{abstract}
\maketitle
The highest energy cosmic rays arrive at Earth with energies in excess of $10^{20}$ eV. Despite decades of work meant to uncover their sources, their origin remains elusive. Observations are limited by the low flux at the end of the cosmic-ray spectrum, calling for the development of new techniques with high duty cycles, high precision, and large surface areas. One promising technique makes a measurement of the radio-frequency electric field from a cosmic-ray air shower, which is nearly linear with the energy of the primary particle. \par
Radio emission arises from a cascade of charges moving inside a dielectric and in the presence of magnetic field in two main ways. Askaryan radiation forms from a charge excess built up in the shower due to Compton, Bhabha, and M{\o}ller scattering and positron absorption, forming a current along the shower axis and radio frequency emission \cite{1962JETP}. Such emission has been measured in accelerator experiments \cite{slac_sand, 2000PhRvE..62.8590G, Gorham:2005cu, 2007PhRvL..99q1101G, 2006PhRvD..74d3002M}. Geomagnetic emission forms when the Lorentz force acts on charges in the shower, generating a time-varying transverse current. The former is present even with no magnetic field, and the latter is present even without a charge asymmetry. In practice, experiments detect the sum of these two effects. \par
Several experiments have detected radio emission from cosmic-ray air showers \cite{1965Natur.205..327J,Porter:1965in,Vernov:1967,Barker:1967ck,Fegan:1968eu,Hazen:1969jd,Hazen:1970cn,SPENCER:1969gb,Fegan:1969fh,Allan:1971,2005Natur.435..313F,2006APh....26..341A,Hoover:2010id,2013A&A...560A..98S,2014PhRvD..89e2002A,Nelles:2015dq}. To progress from event detection to measurement of a differential energy flux, one needs to know the intensity and angular distribution of the radio emission and its frequency dependence. To date, predictions of these parameters have relied on simulations and measurements of air showers themselves, which can be subject to uncertainties in geometry and hadronic interactions in addition to uncertainties related to the radio emission. In contrast, this work provides a direct laboratory benchmark for the simulations of the radio emission using particle showers with well-known cascade physics that develop in a precisely known target and geometry. 
\par
\begin{figure*}
\centering
\includegraphics[width=\textwidth]{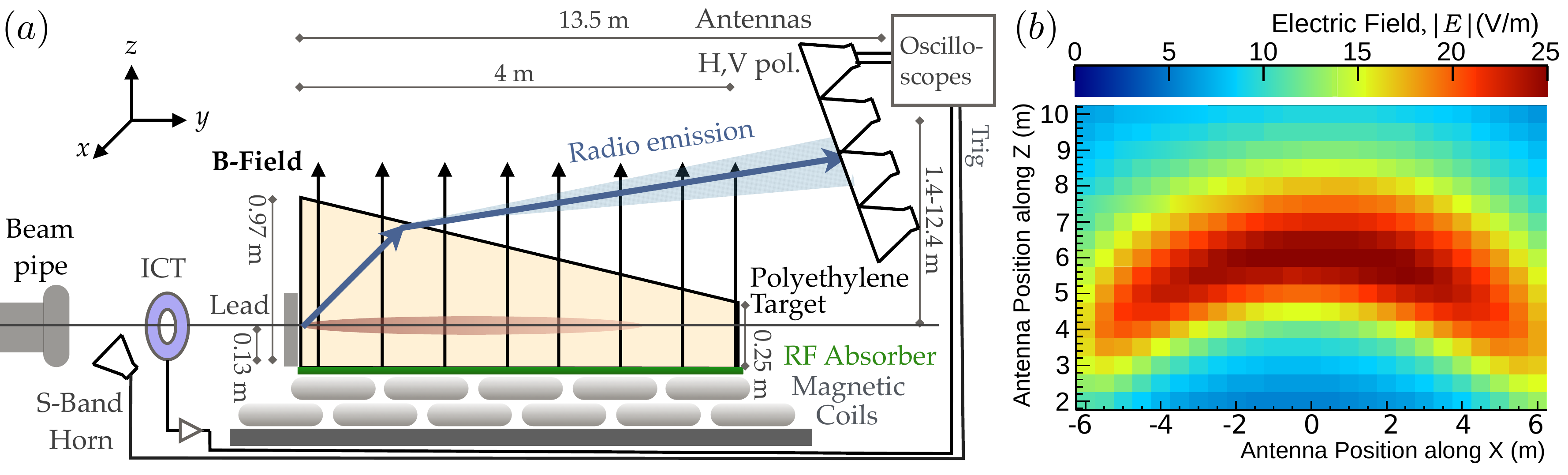}
\caption{(a) Schematic of the experiment, not to scale. (b) Simulated (Endpoints) electric field, $|E|$, in the x-z plane with full magnetic field, 131~pC, and 4.35~GeV. }
\label{fig:sys_diag}
\end{figure*}
Our particle showers developed in a dense plastic instead of rarefied atmosphere. Increasing the density, $\rho$, decreases the shower length, thereby reducing the total amplitude of both radiations integrated over the whole shower. The total Askaryan radiation scales with the shower length as $1/\rho$, weighted by the charge asymmetry. Because the magnetic radiation from a given position in the shower is proportional to the transverse drift velocity of electrons and positrons, the magnetic radiation per unit track length scales with the magnetic field $B$ and the scattering length \cite{Scholten:2008kk}, or $B/\rho$. The total magnetic radiation for the entire shower scales as $B/\rho^2$, meaning that the expected ratio of magnetic to Askaryan radiations scales as $B/\rho$. So, in order to achieve the same relative emission amplitude found in air showers, we increased the magnetic field from the terrestrial 0.5 Gauss to 1000 Gauss, commensurate with an increase in density from air to plastic.\par
Two formalisms, ZHS \cite{1992zhs} and Endpoints \cite{PhysRevE.84.056602}, are at the heart of recent simulations of radio emission, ZHAireS \cite{2012APh....35..325A} and CoREAS \cite{2013AIPC.1535..128H}, respectively. They both treat each shower particle track as an independent radiator, summing up the emission from all tracks in the cascade to obtain the signal that would be received by an observer. The ZHS technique has been adapted to the time-domain \cite{2010PhRvD..81l3009A} and calculates the vector potential of each particle track.
The Endpoints formalism sums the radiation due to the acceleration between discrete segments of the particles' trajectories \cite{PhysRevE.84.056602}. Both formalisms are compared to the accelerator data in this paper.\par
The T-510 experiment, shown schematically in Fig.~\ref{fig:sys_diag}a, took place at the End Station Test Beam (ESTB) in End Station A (ESA) at SLAC. Bunches of electrons with energy 4.35~or 4.55~GeV passed through 2.3 radiation lengths of lead pre-shower and entered a plastic target, generating compact showers with total energy equivalent to a $\sim4\times10^{18}$~eV primary cosmic ray.\par
The target was formed from 5.08~cm~$\times$ 10.16~cm~$\times$ 30.48~cm bricks of high-density polyethylene (HDPE) with density 0.97 g/cm$^3$. Being 4~m long, 0.96~m tall, and 0.60~m wide, it contained the majority of the particle shower. The beam was 0.13~m above the bottom of the target. The bricks on the top surface were machined to a 9.8~degree angle below horizontal to avoid total internal reflection. The index of refraction of HDPE is 1.53, resulting in a refracted Cherenkov angle of 28$^{\circ}$ from the horizontal (49$^{\circ}$ before refraction). Rays emanating from the target at the refracted Cherenkov angle intersected the antenna plane at 6.5~m above the position of the beam. The target floor was lined with an RF absorbing blanket. Several pieces of ANW-79 absorber were placed at both sides of the target and at the exit surface of the target.\par
Fifteen water-cooled coils placed under the target were used to create a vertical magnetic field of up to $\pm$970~G.  This was achieved by supplying sets of five coils in series up to 2400 A direct current with reversible polarity.  The coils were aligned along the beam axis on a 10.16~cm thick steel plate and were staggered in two rows in order to create a more uniform magnetic field. The vertical magnetic field measured at the beam height and maximum current had a RMS variation of 72 G. The magnetic field was strong enough to bring the expected intensity from the magnetic effect to the same order of magnitude as that expected from the Askaryan effect.\par
Four dual-polarization, quad-ridged horn antennas used in the ANITA experiment \cite{Gorham:2009us} recorded the electric fields generated in the particle shower. An overhead crane allowed for the movement of the antenna array to sample the electric field at many positions. The antenna array was placed at the far wall of ESA, 13.5~m from the entrance of the beam to the target. It was tilted at $19.6^{\circ}$ to better align the expected radiation with the boresight of the antennas. The antenna array covered vertical distances between 1.4~m and 12.4~m, corresponding to angles between 40$^{\circ}$ and 55$^{\circ}$ with respect to the beam line at the beam entry point, which, due to the lead pre-shower, is close to shower maximum.\par
The antennas used are sensitive to the 200-1200 MHz band. The comparable frequency range in air showers is lower, because it is inversely proportional to both the Moli\`{e}re radius and $\sin{\theta}$, where $\theta$ is the observation angle. The Moil\`{e}re radius scales as $1/{\rho}$ and $\sin{\theta}$ scales as $\sqrt{\rho}$ to first order. Taken together, we expect that the frequencies scale as $\sqrt{\rho}$. The T-510 bandwidth translates to approximately 10-60 MHz in air showers, comparable to the bandwidth of ground-based air shower experiments. \par
\par
Signals from each antenna ran through 15.24~m of LMR240 coaxial cable and a low-pass filter with a 3~dB point of 1250~MHz to avoid aliasing during data acquisition. Time-series voltages from the horns were collected on 2.0~GHz, 5~GSa/s oscilloscopes. A global trigger was provided by the broadband transition radiation produced by the beam exiting the beam pipe, collected in an S-band horn antenna. Events were recorded at 1~Hz.\par
The beam charge was measured using an integrating charge transformer situated between the beam pipe and the target. The mean bunch charge was 131~pC, with a shot-to-shot standard deviation of 3~pC. Measurements of the bunch charge at several positions indicate a 2\% systematic uncertainty. \par
Particle showers were simulated with GEANT4, using the measured magnetic field. Five thousand primary electrons were injected, and the results were scaled to 131~pC. The radio emission was then simulated following the ZHS and Endpoints formalisms. In calculating the radiation from each track we included refraction, Fresnel coefficients, and demagnification effects~\cite{2004PhRvD..69a3008L} at the surface of the target.\par
 The simulation, shown on the right in Fig. \ref{fig:sys_diag}b, demonstrates that the expected radiation forms a ring when projected onto a two-dimensional plane 13.5 m from the entrance to the target, peaking at about 6.5 m above the shower axis. The electric field strength, $|E|$, map shows the superposition of magnetic and Askaryan components. Since the magnetic field is vertical at the shower, the former is horizontally polarized, whereas the polarization of the Askaryan contribution points radially from the shower axis. The interference between the two produces the left-right asymmetry shown in the figure. Refraction at the target surface makes the ring elliptical rather than circular. The ring is cut off on both sides due to the finite target. The simulation is done using ray optics. The top of the target acts as a diffractive slit, with a Fresnel zone of about 60 cm at 300 MHz, which is smaller than the length of the target. The target width ($\pm 30$~cm) corresponds to a phase lag of about $30^\circ$, and so we expect the simulation to modestly overestimate $|E|$ at low frequency. Reflections were not included in Fig.~\ref{fig:sys_diag}b, but are discussed below. Effects due to transition radiation were estimated and found to be two orders of magnitude below the Askaryan radiation. \par
\begin{figure}
\centering
\includegraphics[width=\columnwidth]{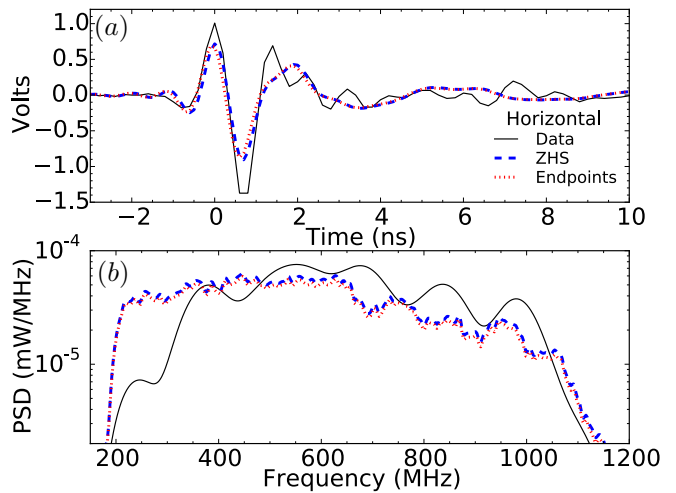}	
\caption{(a) Simulated and measured voltages 6.52~m above the beam in the horizontal polarization channel at full magnetic field. (b) Corresponding power spectral densities.}
\label{fig:wave}
\end{figure}
By design, at our antenna locations, the two types of radiation separate into orthogonal polarizations. The magnetic is horizontally polarized, while the Askaryan is vertical. However, in real antennas, the vertical signal leaks into the horizontally polarized channel at about the 25\% level in amplitude (about $-12$~dB in power).  We eliminate this leakage in the horizontal polarization by construction in using the difference between field-up, $V_{B_+}(t)$, and field-down, $V_{B_-}(t)$, data, namely $V(t)=\frac{1}{2}( V_{B_+}(t) - V_{B-}(t))$. Each waveform is also scaled by the beam bunch charge to 131~pC. When this construction is applied to the vertically polarized signal, we see only noise, as expected. The resulting single-acquisition waveform from the horizontal polarization at maximum magnetic field strength is shown in Fig.~\ref{fig:wave}a. Fig.~\ref{fig:wave}b shows the power spectral density normalized by the sampling interval 200 ps/point for the data and 100 ps/point for the simulations.\par
For comparison with the data, the simulated electric fields were convolved with the measured antenna effective heights \cite{Gorham:2009us} and response due to the cables and filters used in the system. The convolution was performed in the frequency domain following standard techniques \cite{Krauss:1988bx}. The predicted values in Fig.~\ref{fig:wave} from the ZHS and Endpoints formalisms agree to within 3\% in peak amplitude and 7\% in integrated power. The dominant features in the time-domain waveform arise from antenna response and filters, which the two simulations have in common. The shape of the simulated waveforms reproduces the data well, giving us confidence in the experimental modeling. The absolute scale is discussed below.\par
Internal reflections from the bottom of the target interfered with the signal transmitted through the top surface of the target, which is apparent in the modulation of the power spectral density in Fig.~\ref{fig:wave}b. A low amplitude reflection with two internal bounces arrived at the antenna $\sim 7$~ns after the main pulse, and was responsible for the $\Delta f \simeq 150$~MHz frequency beat in Fig. \ref{fig:wave}b. More important is the radiation that reflected only once off the bottom of the target. Assuming that the absorber has a higher index of refraction than the HDPE, that reflection arrived at the antenna $\sim 1$~ns after the direct pulse, but with inverted polarity. Adding a time delayed equal amplitude reflection to the direct pulse for a particular antenna varies the simulated peak time-domain amplitude by a factor of up to 1.38 for horizontal polarization and 1.43 for vertical polarization. Thus, averaged over 300-1200 MHz, the uncertainty in the models due to reflections is of order 40\% for both polarizations. We exclude the 200-300 MHz band in the data and simulation comparisons due to uncertainties related to diffraction and in the antenna response at low frequencies \cite{Gorham:2009us}. For the horizontal polarization shown in Fig. \ref{fig:wave}a, the time-domain peaks of the data exceed the simulations by 35\%, commensurate with the systematic uncertainty.\par
\begin{figure}
\centering
\includegraphics[width=\columnwidth]{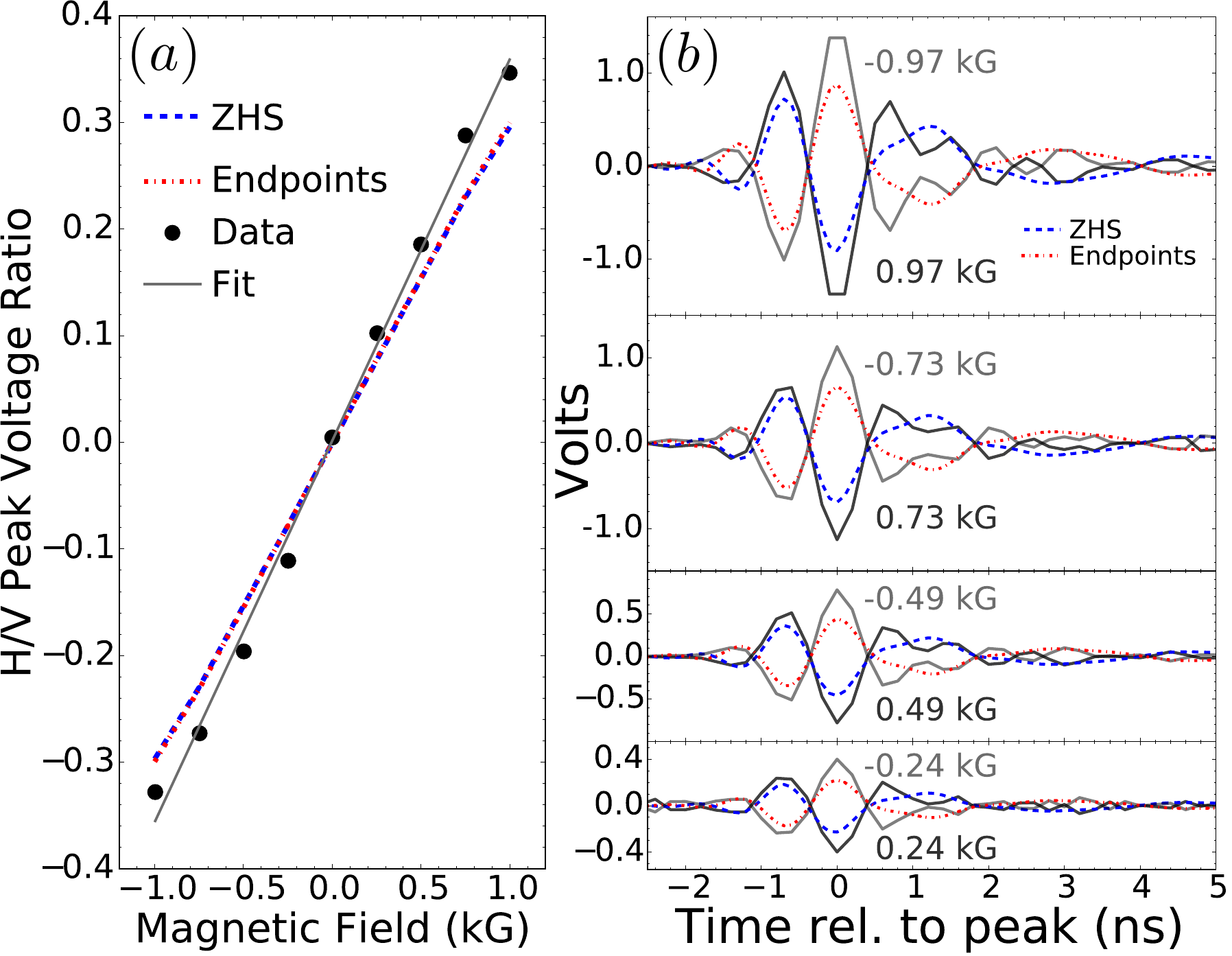}
\caption{(a) Horizontally polarized signal normalized by vertical showing the expected linear behavior vs. magnetic field. (b) The oscilloscope traces (solid) show the polarity flip. Models (dashed) are shown for opposite polarities.
} 
\label{fig:ramp}
\end{figure}
Fig.~\ref{fig:ramp} shows that the amplitude of the horizontally polarized emission is linearly dependent on the magnetic field. The polarity of this induced voltage changes sign when the direction of the magnetic field flips direction, indicating that the transverse current flows in the opposite direction. The vertically polarized emission is observed to be constant with respect to magnetic field strength.  The difference in slopes between the data and simulation is 20\% which given our current systematic uncertainty should be taken as agreement between the two. This agreement along with the expectation that the ratio of magnetic emission to Askaryan emission scales with $B/\rho$ confirms that transverse currents generate magnetic emission in air showers.\par
\begin{figure}
\centering
\includegraphics[width=\columnwidth]{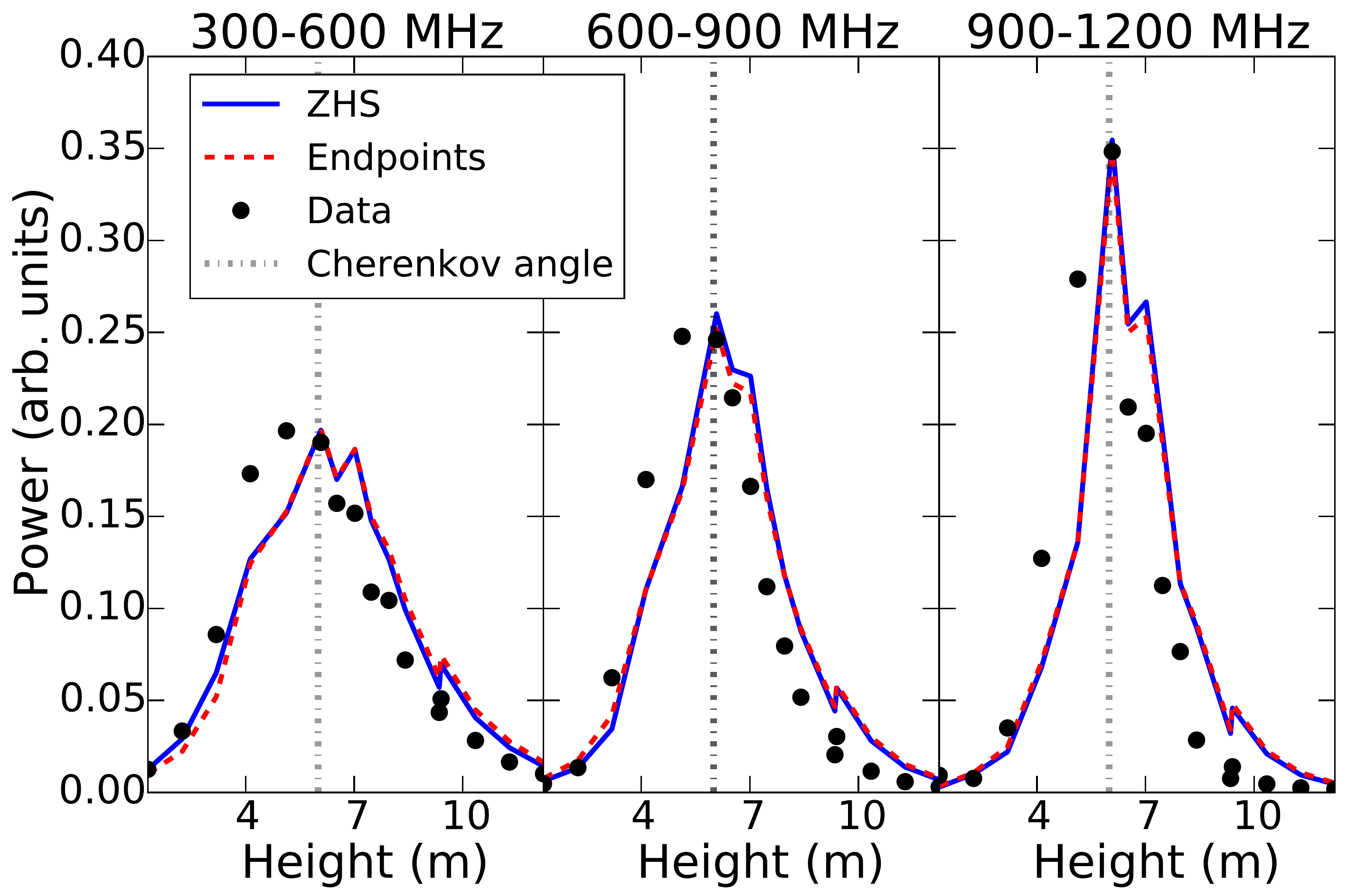}
\caption{Beam patterns for three frequency bands in the horizontal polarization at 970~G.}
\label{fig:cone}
\end{figure}

Several aspects of the radio emission in cosmic-ray air showers contribute to the formation of a conical beam pattern centered around the Cherenkov angle \cite{2013AIPC.1535..209B,2011PhRvL.107f1101D}. In this experiment, the angular radiation pattern was measured by placing the antennas at different vertical positions. The power profile, which traverses the expected peak of the cone, is shown in Fig.~\ref{fig:cone} for three different frequency bands. Each profile is normalized by its total power.\par
The observed cones are somewhat different than those from air showers. In both cases, the expected power spectrum observed at the Cherenkov angle peaks at a frequency determined by the transverse size of the shower $L_\perp$, which is a fraction of a radiation length $X_0$. The Cherenkov cones have widths $\delta \theta = c \phi/(n f L_\parallel \tan \theta_C)$, determined by the angle over which the shower is coherent, $\phi$, the frequency of observation, $f$, the Cherenkov angle, $\theta_{C}$, and the shower length, $L_{\parallel}$. At the peak power frequency, the width of the Cherenkov cone is determined simply by the aspect ratio of the shower $\delta \theta = L_\perp/L_\parallel$. For the T-510 beam of 4.5 GeV electrons, $L_\parallel \simeq 2 X_0$ and we both predict and observe $\delta\theta \simeq 5^\circ$. For air showers induced by $10^{17}$~eV primaries, $L_\parallel \simeq 5 X_0$, and the inner edge of the Cherenkov cone is washed out because $\theta_C (1^{\circ}) < \delta\theta (2^{\circ})$, causing the Cherenkov feature to appear as a filled-in disk, as observed by LOPES \cite{Apel:2014bk}. At higher frequencies, the width of the Cherenkov ring scales as $1/f$, and the ring becomes well defined, as observed at LOFAR \cite{Nelles:2015dq} and inferred by ANITA \cite{2015arXiv150605396S}. The accurate simulation of the Cherenkov cone and frequency behavior shown in Fig. \ref{fig:cone} is directly relevant to the ability of the simulations to model emission from air showers.\par
%
The signal polarization observed in T-510 confirms the paradigm that transverse currents due to the geomagnetic effect and longitudinal currents due to charge excess produce the radiation observed in air showers, consistent with measurements by CODALEMA \cite{Belletoile201550} and  AERA \cite{2014PhRvD..89e2002A}. The observed Cherenkov cone in this work and at LOFAR \cite{Nelles:2015dq} indicate that the refractive index is an important component to accurate modeling of the electrodynamics. Recent comparisons between the microscopic calculations and LOPES data confirm that first-principle calculations accurately predict the absolute scale of the radio emission \cite{2016APh....75...72A}, but the measurement is subject to uncertainties associated with air shower physics such as composition and hadronic interaction models. In using a fixed beam geometry and electromagnetic shower composition, this experiment confirms the absolute scale of the microscopic calculations with different systematic uncertainties.\par
We have presented the first laboratory benchmark of radio-frequency radiation from electromagnetic cascades under the influence of a magnetic field. We compared the radio emission produced in a well-defined target geometry with a well-defined particle shower to predictions made by microscopic models, which rely on first principles of electrodynamics and have no free parameters. The models agree and accurately predict the absolute scale of the radio emission to within our systematic uncertainty. The observed radiation grows linearly with magnetic field strength and forms a conical beam centered around the Cherenkov angle. Being in a relevant frequency range and independent of hadronic interaction models and complex geometries, this experiment is complementary to \textit{in situ} observations of radio emission from air showers by current experiments \cite{Gorham:2009us,2015arXiv150605396S, 2013A&A...560A..98S,Nelles:2015dq, 2006APh....26..341A, 2014PhRvD..89e2002A}. It also strengthens the case for proposed experiments based on the radio technique \cite{2011APh....35..242G, 2013arXiv1302.1263R, 2015Taroge, 2015GRAND}. T-510 provides strong evidence that the electromagnetic simulations can be used to reliably predict the radio emission from extensive air showers.\par

\begin{acknowledgments}
We thank the referees and editors for critical remarks leading to an improved presentation. The authors thank SLAC National Accelerator Laboratory for providing facilities and support and especially Janice Nelson and Carl Hudspeth for their support and dedication that made T-510 possible. We thank D.~Z.~Besson for helpful discussions. This material is based upon work supported by the Department of Energy under Award Numbers DE-AC02-76SF00515, \text{DE-SC0009937}, and others. Work supported in part by grants from the National Aeronautics and Space Administration and the Taiwan Ministry of Science and Technology under project number MOST103-2119-M-002-002, among others. Part of this research was  funded through the JPL Internal Research and Technology Development program. This work was supported in part by the Kavli Institute for Cosmological Physics at the University of Chicago through grant NSF PHY-1125897 and an endowment from the Kavli Foundation and its founder Fred Kavli. K. Belov acknowledges support from the Karlsruher Institut f\"ur Technologie under a guest fellowship. We are grateful to the ANITA collaboration for use of antennas and other equipment. 
\end{acknowledgments}

\bibliography{t-510}
\clearpage


\section*{Supplementary material}

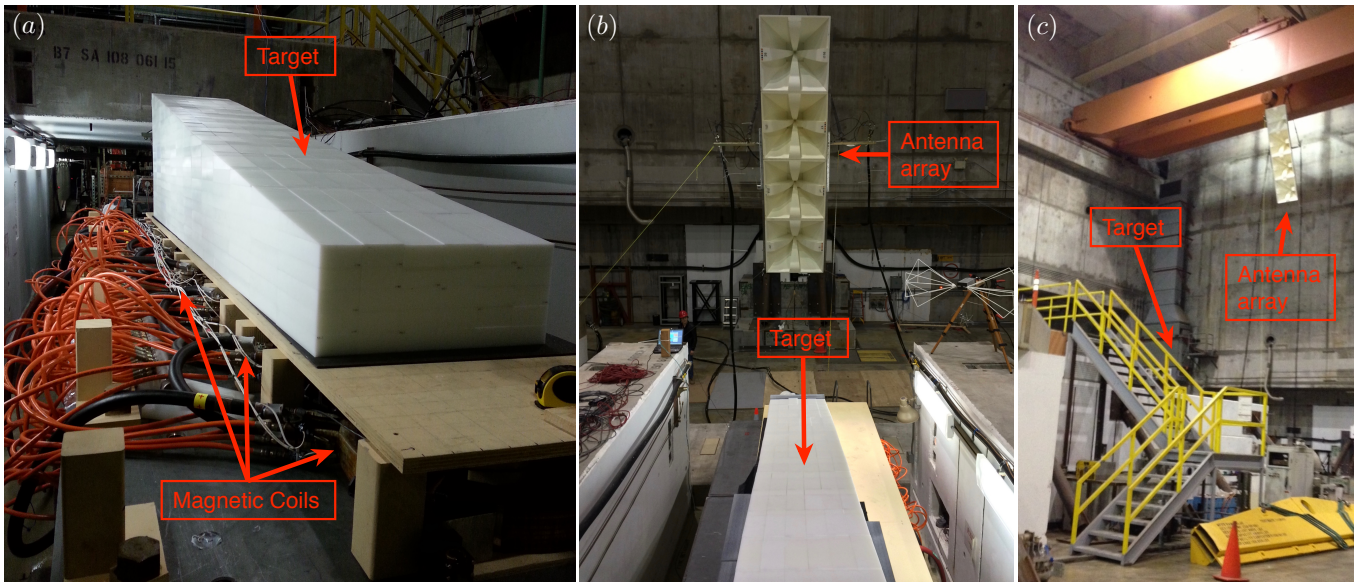

FIG. S1: (a) The HDPE target stacked on top of the magnetic field coils. The quad-ridged horn array in the $y$-$z$ plane (b) shown close to the target for scale and (c) in a position for data-taking (12.4 m above the beam, and 13.5 m from the beam entrance to the target.)

# Supplemental Material: Accelerator measurements of magnetically-induced radio emission from particle cascades with applications to cosmic-ray air showers

## I. THE T-510 EXPERIMENTAL DESIGN

The T-510 experiment was designed to call out the critical features of geo-magnetic radiation. To do this, we used a geometry that separated the magnetic radiation and the Askaryan radiation into two different polarization states. Fig. S1 and Fig. 1a show the experimental setup, where an HDPE target was constructed on top of a series of magnetic field coils. Radiation produced in the target was recorded by an antenna array shown in the near field in Fig. S1b. For the data presented in this work, the antenna array was further away (13.5 m) from the entrance of the beam to the target, as shown in Fig. S1c.

The magnetic field map at maximum current is shown in Fig. S2, measured at the beam height, 21 cm above the coils with 5 cm × 5 cm grid spacing and 3.64 G precision. There is a 6% uncertainty on the magnetic field measurements arising from the average percent error of the Hall probe used. Because the vertical ($B_z$) field falls off near the edges of the coils, the edge of the target was placed in the middle of the first coil. Along the beam position, the average magnetic field in the vertical direction is 845 G, and the RMS variation is 72 G. The peak magnetic field strength is 970 G. In our simulations of the T-510 experiment, we included the measured, three-dimensional magnetic field.

As shown in Fig. S2, the magnetic field along the beam is strongest in the vertical ($z$) direction. Because of this, the transverse current that develops in the shower is primarily oriented along the transverse ($x$) axis, thereby producing horizontally-polarized radiation. In contrast, the Askaryan radiation that arises from the longitudinal current in the shower is radially polarized. When the antenna array is confined to the line defined by the beam and the vertical direction, the magnetic radiation observed at the antennas is horizontally polarized, while the Askaryan radiation is vertically polarized. While the ZHS and Endpoints models calculate the radiation per particle track and are therefore ambivalent to macroscopic effects, the simulation shown in Fig. 1b confirms this simplified picture.

## II. SCALING TO COSMIC RAY AIR SHOWERS

The goal of the T-510 was not to accurately replicate all the conditions of an air shower in the lab. Rather, the goal was to create a model system with relative field strengths arising from both magnetic and Askaryan emission comparable to those found in air showers. Even though direct comparisons between cosmic ray air showers and the

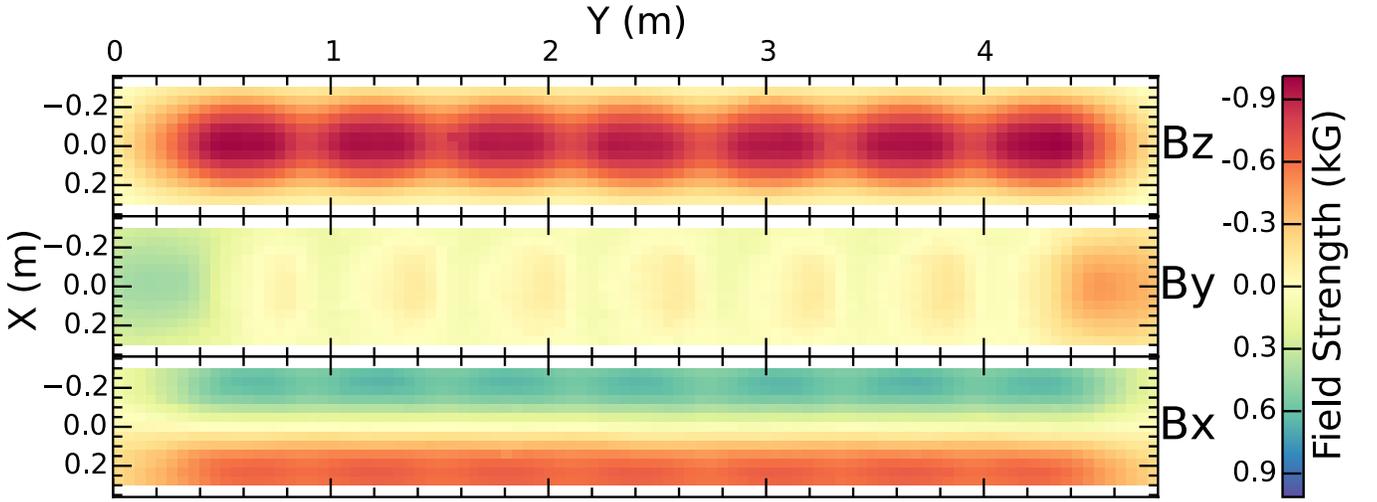

FIG. S2: The measured magnetic field at the beam height along the vertical ($z$), longitudinal ($y$), and transverse ($x$) axes.

showers in T-510 are difficult, the radiations observed in T-510 are relevant for air shower experiments, because the densities, frequencies of observation, and the magnetic field strength are scaled appropriately.

In air showers the geomagnetic emission dominates over Askaryan emission, but in a dense dielectric such as HDPE, the Askaryan radiation dominates because it scales with the shower length. The source for both radiations is the current distribution in the shower. For Askaryan radiation, the total longitudinal current depends on the total track length of particles, weighted by the charge asymmetry. The total track length scales as the radiation length or as $1/\rho$. For the magnetic radiation, the total transverse current, $J_\perp$ depends on the total track length weighted by the transverse velocity, or the accumulated transverse acceleration, $a_\perp$, during a radiation length. $J_\perp \sim a_\perp t$ over time $t$. The acceleration increases with $B$, and the time of acceleration increases as the radiation length or $1/\rho$. The total transverse current therefore scales as $B/\rho^2$. To recreate a comparable ratio of source currents for magnetic emission to Askaryan emission as observed in an air shower, we preserved the ratio $(B/\rho^2)/(1/\rho) = B/\rho$. To mimic the relative importance of geomagnetic radiation to Askaryan radiation in air showers, the quantity $B/\rho$ in T-510 is comparable to that in air showers.

Two delay times, the transverse delay time $\delta t_1$ and the longitudinal delay time $\delta t_2$, determine the frequencies and angles over which showers are coherent. The transverse delay time determines the range of frequencies where the radiation is in phase, while both the longitudinal and transverse set the range of angles over which the radiation is coherent. The bandwidth observed in T-510 is higher than the corresponding bandwidth for air shower experiments. The shape of the observed beam pattern in depends on both the Cherenkov angle and beam width, both of which differ in T-510 and in air shower experiments.

We first assume that the radiation is coherent over the transverse dimensions of the shower, $L_\perp$, which we take to be one twelfth the Molière radius, $R_m$. The observation frequency, $f$, is comparable to the time difference, $\delta t_1$, between radiation from the center of the shower (point $a$) and from the edge of the shower (point $b$) at the observation angle, $\theta$, as shown in Fig. S3. The transverse delay time is then given by

$$\delta t_1 = \frac{nL_\perp \sin\theta}{c} \tag{S1}$$

where $\theta$ is the observation angle. At the Cherenkov angle, $\theta_C$, $f \sim 1/\delta t_1 \sim c/(L_\perp \sin\theta_C) \sim c/(R_m \sin\theta_C)$. In T-510, $\sin\theta_C = 0.8$, but in air showers $\sin\theta_C$ scales as $1/\sqrt{\rho}$. Therefore, the T-510 measurements made at 200-1200 MHz are comparable with observations of air showers in the 10-60 MHz band.

Assuming that the angle $\theta$ in Fig. S3 is given by $\theta_C + \delta\theta$, the power spectrum at $\theta_C$ peaks at a frequency $f_0 = \phi/\delta t_1$, corresponding to the highest frequency over which the shower is coherent. On the Cherenkov cone, $\delta\theta = 0$, such that:

$$f_0 = \frac{c\phi}{nL_\perp \sin\theta_C} = \frac{c\phi}{L_\perp \tan\theta_C} \tag{S2}$$

The longitudinal delay time is the delay time across the length of the shower, or from point $c$ relative to point $a$ in Fig. S3. It includes the time required for the particle shower to propagate from $a$ to $c$ and the delay time between radiation emitted at point $a$ and point $c$.

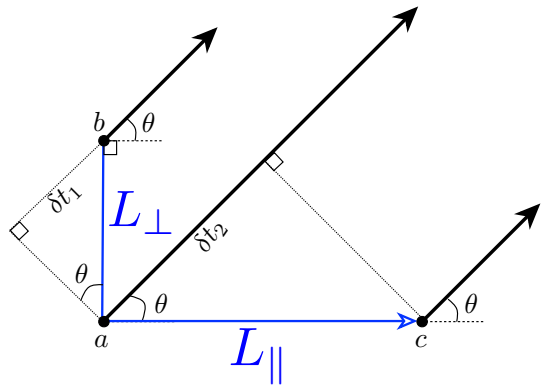

FIG. S3: Schematic of transverse ($\delta t_1$) and longitudinal ($\delta t_2$) coherence times.

$$\delta t_2 = \frac{L_\parallel}{c} - \frac{nL_\parallel}{c}\cos\theta = \frac{L_\parallel}{c}(1 - n\cos\theta) \tag{S3}$$

In solid media, the width of the Cherenkov cone is small: $\delta\theta << \theta_C$.

$$\cos\theta \simeq \cos\theta_C - \delta\theta\sin\theta_C \tag{S4}$$

$$\delta t_2 \simeq \frac{nL_\parallel}{c}\delta\theta\tan\theta_C \tag{S5}$$

Again, assuming the angle over which the radiation is coherent is $\phi = f\delta t_2 = \frac{nL_\parallel}{c}f\delta\theta\tan\theta_C$, then the width of the Cherenkov cone is:

$$\delta\theta = \frac{c\phi}{fnL_\parallel\tan\theta_C} = \frac{f_0}{f}\frac{c\phi}{f_0 nL_\parallel\tan\theta_C} = \frac{f_0}{f}\frac{L_\perp}{L_\parallel} \tag{S6}$$

At $f = f_0$, the width of the cone is determined by the shape of the shower. The width is narrower at higher frequencies and broader at lower frequencies. From the simulations, the showers in T-510 are two radiation lengths long and the cutoff frequency is approximately 1 GHz, and so $\delta\theta \sim 5°$, similar to the observed widths in Fig. 7. The beams observed in T-510 are narrow cones, because $\theta_C > \delta\theta$.

Finally, we note that the density of the atmosphere is changing in air showers, while it is fixed for T-510. This is important when the the radiation length becomes comparable to the atmospheric scale height, which does occur for highly inclined showers such as those observed by ANITA [S1] and proposed sub-orbital experiments [S2, S3]. However, for the less-inclined showers observed by ground arrays, the density at the point of emission rather than the density gradient determines the radiation strength.

---


\end{document}